# Investigation on the superluminality of evanescent modes via quantum Lorentz transformation

**Zhi-Yong Wang***, **Cai-Dong Xiong**

School of Optoelectronic Information, University of Electronic Science and Technology

of China, Chengdu 610054, CHINA

*E-mail: zywang@uestc.edu.cn

**Abstract**

Applying the fact that guided photons inside a waveguide can be treated as massive particles, one can study the superluminality of evanescent modes via showing that a massive particle can propagate over a spacelike interval, which corresponds to quantum tunneling effects. For this purpose, we treat the particle as a quantum reference frame, while attach an inertia observer to a classical reference frame, and then quantize the formulae for the Lorentz transformation between the quantum and classical reference frames, from which we obtain the conclusion that, owing to the Heisenberg's uncertainty relation, the particle can propagate over a spacelike interval.

**1. Introduction**

Recently we have presented theoretical evidence, based on quantum field theory, for the superluminality of evanescent modes in undersized waveguides [1-3]. To make our conclusion more convincing, we will present another theoretical evidence for such superluminality by reformulating special relativity on the basis of quantum mechanics, and from which we will obtain the same conclusion.

According to special relativity, a particle cannot propagate over a spacelike interval. As a result, in spite of the fact that such superluminal behavior does actually exist according to quantum field theory [1-5], there have been many controversies about the existence of this kind of superluminal phenomenon. For example, though both theoretical and experimental studies had obtained the same conclusion that photons inside an undersized waveguide propagate superluminally [6-12], many papers disproving this conclusion have been published [13-18] by Winful *et al*. After all, to their mind, particle's superluminal propagation does not conform to special relativity, and it is impossible for special relativity to contain errors. However, special relativity has been developed on the basis of classical mechanics without taking into account any quantum-mechanical effect, which implies that some traditional conclusions in special relativity might be modified on condition that quantum-mechanical effects cannot be ignored. Therefore, a convincing argument for the superluminality of evanescent modes should be obtained by reformulating special relativity basing on quantum mechanics.

In fact, one can combine special relativity with quantum mechanics via two different approaches: (1) developing quantum mechanics on the basis of special relativity, one can obtain relativistic quantum theory (including relativistic quantum mechanics and quantum field theory); (2) developing special relativity on the basis of quantum mechanics, one might obtain a quantum-mechanical special relativity. The former has been successful, while the latter remains to be achieved. Historically, many attempts have been made to investigate how quantum effects might modify special relativity (e.g., try to apply quantum-mechanical uncertainty to the reference frames of relativity; try to extend the concept of macroscopic observers to include that of quantum observers, etc.) [20-22], a quantum reference frame defined by a material object subject to the laws of quantum mechanics has been studied [23-27]. However, these attempts have not been completely successful. For example, in Ref. [23] quantum reference frame has been discussed within the framework of nonrelativistic quantum theory, such that it has been concerned with Galilean relativity, instead of Einstein relativity. Furthermore, to take a "quantum special relativity" as being a limit of quantum gravity in a similar way Special Relativity is a limit of General Relativity, Doubly Special Relativity has been proposed [28-31], whose idea is that there exist in nature two observer-independent scales, of velocity, identified with the speed of light, and of mass, which is expected to be of order of Planck mass. However, even if Doubly Special Relativity is valid, it does not deviate from the usual Special Relativity unless the scale under consideration approaches the Planck scale, and thus it has nothing to do with our present issue.

Any way, a proper quantum-mechanical special relativity should give us some same conclusions as those from relativistic quantum theory (such as quantum field theory). For our purpose, we will study the Lorentz transformation between classical and quantum reference frames, and quantizes its classical expression, from which we can obtain the same conclusion as that from quantum field theory: a particle can propagate over a spacelike interval. Because guided photons inside a hollow waveguide can be treated as free massive particles, this effort can provide another theoretical evidence for the superluminality of evanescent modes.

**2. Quantum Lorentz transformation**

Consider two inertial reference frames $S$ and $S'$ with a relative velocity

$v = (v, 0, 0)$ between them. We shall denote observables by unprimed variables when referring to $S$, and by primed variables when referring to $S'$, and then the time and space coordinates of a point are denoted as $(t, x, y, z)$ and $(t', x', y', z')$ in the frames $S$ and $S'$, respectively. The coordinate axes in the two frames are parallel and oriented so that the frame $S'$ is moving in the positive x direction with speed $v > 0$, as viewed from $S$. Let the origins of the coordinates in $S$ and $S'$ be coincident at time $t = t' = 0$. All statements here are presented from the point of view of classical mechanics, or, in other words, they are valid in the sense of quantum-mechanical average.

From the physical point of view, a frame of reference is defined by a material object of the same nature as the objects that form the system under investigation and the measuring instruments [24]. If the mass of the material object is finite, the corresponding reference frame (say, quantum reference frame) would be subject to the laws of quantum mechanics, and the interaction between object and measuring device might not be neglected. In particular, Heisenberg's uncertainty relations forbid the exact determination of the relative position and velocity of quantum reference frame. For simplicity, we assume that the interaction between a physical system and measuring device is so small that all quantum reference frames can approximatively be regarded as inertial ones (they are inertial ones in the sense of quantum-mechanical average).

To study whether a particle can propagate over a spacelike interval, we assume that the frame $S'$ is attached to a particle Q with rest mass *m* (i.e., a quantum-mechanical object of finite mass), such that the frame $S'$ can be regarded as consisting of a measuring device and the particle Q. For simplicity, we assume that the mass of the measuring device can be ignored as compared with that of the particle Q. As a result, the frame $S'$ can approximatively be defined by the particle Q with rest mass $m$, wherein a Cartesian coordinate system is chosen in such a manner that the coordinates of the particle Q is $(t', x', 0, 0)$ as viewed in $S'$, and is $(t, x, 0, 0)$ as viewed in the frame $S$.

On the other hand, for convenience we assume that the frame $S$ has an infinite mass. In other words, the frame $S$ is a classical reference frame while the frame $S'$ is a quantum one. For simplicity, from now on we will omit the y- and z- axes. According to the Lorentz transformation one has

$$\begin{cases} x' = (x - vt) / \sqrt{1 - (v^2/c^2)} \\ t' = [t - (vx/c^2)] / \sqrt{1 - (v^2/c^2)} \end{cases}, \quad (1)$$

Because the frame $S'$ is attached to the particle Q, let $\boldsymbol{p} = (p, 0, 0)$ and $E$ denote the momentum and energy of the particle Q as observed in the frame $S$, respectively, then $p = Ev/c^2 > 0$. In other words, as observed in $S$, the particle Q has the 4D momentum $(E, p, 0, 0)$ and the 4D coordinate $(t, x, 0, 0)$. Using $E^2 = p^2 c^2 + m^2 c^4$ and $v = pc^2/E$, Eq. (1) can be rewritten as

$$\begin{cases} x' = (Ex - c^2 pt) / mc^2 \\ t' = (Et - px) / mc^2 \end{cases}. \quad (2)$$

As we know, the transition from the classical expression (2) to a quantum-mechanical one requires us to symmetrize Eq. (2) and replace all its variables with the corresponding operators, in such a way we formally give a quantum Lorentz transformation (in the position-space representation)

$$\begin{cases} x' = [(\hat{H}x + x\hat{H}) - c^2(\hat{p}t + t\hat{p})] / 2mc^2 \\ t' = [(\hat{H}t + t\hat{H}) - (\hat{p}x + x\hat{p})] / 2mc^2 \end{cases}. \quad (3)$$

where $\hat{H}$ is the Hamilton operator satisfying $\hat{H}^2 = \hat{p}^2 c^2 + m^2 c^4$ and $\hat{p} = -i\hbar \partial/\partial x$ ($\hbar$ is the Planck constant divided by $2\pi$). Using $dt/dt = \partial t/\partial t + (i/\hbar)[\hat{H},t] = \partial t/\partial t = 1$ one has

$$[\hat{H},t] = \hat{H}t - t\hat{H} = 0. \tag{4}$$

That is, in contrast with the conjugate pair $x$ and $\hat{p} = -i\hbar\partial/\partial x$, $\hat{H}$ and $t$ do not constitute a conjugate pair. Likewise, owing to $\partial t/\partial x = 0$, one has $t\hat{p} = \hat{p}t$. Therefore, as viewed in the classical reference frame $S$, time coordinate $t$ acts as a parameter rather than an operator, which is in agreement with the traditional conclusion (as a result, time in quantum mechanics has been a controversial issue since the advent of quantum theory). Using $\hat{H}t = t\hat{H}$ and $t\hat{p} = \hat{p}t$ the quantum Lorentz transformation (3) can be rewritten as

$$\begin{cases} x' = \dfrac{(\hat{H}x + x\hat{H})}{2mc^2} - \dfrac{t\hat{p}}{m} \\ t' = \dfrac{t\hat{H}}{mc^2} - \dfrac{(\hat{p}x + x\hat{p})}{2mc^2} \end{cases}. \tag{5}$$

Consider that the particle Q moves relative to the frame $S$ with constant velocity $v$ along x-axis, one has $dv/dt = d^2x/dt^2 = 0 = (i/\hbar)[\hat{H},(i/\hbar)[\hat{H},x]]$, i.e., $\hat{H}[\hat{H},x] = [\hat{H},x]\hat{H}$, it follows that

$$[\hat{H}^2, x] = \hat{H}[\hat{H},x] + [\hat{H},x]\hat{H} = 2\hat{H}[\hat{H},x]. \tag{6}$$

. On the other hand, using $\hat{H}^2 = \hat{p}^2c^2 + m^2c^4$ one has

$$[\hat{H}^2, x] = \hat{p}[\hat{p},x]c^2 + [\hat{p},x]\hat{p}c^2 = -2i\hbar\hat{p}c^2. \tag{7}$$

Combining Eq. (6) with Eq. (7), one has:

$$[\hat{H}, x] = -i\hbar\hat{H}^{-1}\hat{p}c^2. \tag{8}$$

From Eq. (8) one can obtain the desired result $dx/dt = (i/\hbar)[\hat{H},x] = \hat{H}^{-1}\hat{p}c^2$, which is related to the classical expression $v = dx/dt = pc^2/E$ and in agreement with Ehrenfest's theorems. In fact, take Dirac electron for example, by splitting up every operator into an even and an odd part so as to throw off the zitterbewegung part [32], one can obtain a *true* velocity operator that is similar to $dx/dt = \hat{H}^{-1}\hat{p}c^2$.

**3. Existence of particle superluminal propagation**

Applying $\hat{H}t = t\hat{H}$, $t\hat{p} = \hat{p}t$, $\hat{p}x = x\hat{p} - i\hbar$, $\hat{H}x = x\hat{H} - i\hbar\hat{H}^{-1}\hat{p}c^2$, $\hat{H}^2 = \hat{p}^2c^2 + m^2c^4$, $\hat{H}\hat{p} = \hat{p}\hat{H}$, $xt = tx$, and Eq. (8), one can obtain (see **Appendix A**):

$$c^2t'^2 - x'^2 = c^2t^2 - x^2 + \hbar^2c^2\hat{H}^{-2}/4. \tag{9}$$

Owing to $\hat{H}^2 = \hat{p}^2c^2 + m^2c^4 \geq m^2c^4$ (in the sense of eigenvalues or quantum-mechanical averages of operators), for a timelike or lightlike interval $c^2t'^2 - x'^2 \geq 0$, using Eq. (9) one has

$$c^2t^2 - x^2 \geq -\hbar^2c^2\hat{H}^{-2}/4 \geq -\hbar^2/4m^2c^2 = -(\lambdabar/2)^2, \tag{10}$$

where $\lambdabar = \hbar/mc$ is the Compton wavelength of the particle Q. Eq. (10) implies that, as observed in $S$, the particle Q can propagate over a spacelike interval provided that

$$0 > c^2t^2 - x^2 \geq -(\lambdabar/2)^2, \tag{11}$$

which is in agreement with the traditional conclusion derived from quantum field theory [4, 33] (as discussed later). Moreover, via Refs. [1-4] one can show that the particle Q satisfying Eq. (11) corresponds to the one tunneling through a potential barrier (including photons tunneling through an undersized waveguide, as discussed later).

On the other hand, let $\Gamma(t,x)$ denote the probability amplitude for the particle Q to propagate from $(0,0)$ to $(t,x)$, according to Ref. [4], for spacelike interval $c^2t^2 - x^2 < 0$ (for the moment the particle Q tunneling through a potential barrier), one has:

$$\Gamma(t,x) \propto \exp(-\sqrt{x^2 - c^2t^2}/\lambdabar). \tag{12}$$

Therefore, as for the probability (say, $P(t,x)$) for the particle Q to propagate from $(0,0)$ to $(t,x)$, one has

$$P(t,x) \propto |\Gamma(t,x)|^2 \propto \exp(-2\sqrt{x^2-c^2t^2}/\lambdabar). \tag{13}$$

In numerical analysis, one always takes $P(t,x) \propto \exp(-2\sqrt{x^2-c^2t^2}/\lambdabar) \geq \exp(-1) \approx 0$, by which a characteristic length of $\lambdabar/2$ is defined, and an effective range $0 < \sqrt{x^2-c^2t^2} < \lambdabar/2$ of the spacelike interval is determined. In other words, the probability $P(t,x)$ for the particle Q to superluminally propagate from $(0,0)$ to $(t,x)$ CANNOT be ignored provided that $0 > c^2t^2 - x^2 \geq -(\lambdabar/2)^2$, which is in agreement with Eq. (11). Because Eq. (11) is based on quantum mechanics (the first- quantized theory) while Eq. (13) is based on quantum field theory (the second- quantized theory), Eq. (11) corresponds to an approximation of Eq. (13), i.e., the numerical approximation $P(t,x) \propto \exp(-2\sqrt{x^2-c^2t^2}/\lambdabar_c) \geq \exp(-1) \approx 0$.

By the way, taking $\Gamma(t,x) \propto \exp(-\sqrt{x^2-c^2t^2}/\lambdabar) \geq \exp(-1) \approx 0$, one has $0 > c^2t^2 - x^2 \geq -\lambdabar^2$, which is the same as that presented in Ref. [33]. However, the probability $P(t,x) \propto |\Gamma(t,x)|^2$ is an observable while the probability amplitude $\Gamma(t,x)$ is not. Therefore, the result presented in Ref. [33] is just a rough estimate.

**4. Superluminality of evanescent modes**

Let us consider guided photons inside a hollow metallic waveguide being placed along the direction of x-axis. In Ref. [1] we have shown that the behaviors of guided waves are the same as those of de Broglie matter waves (in Ref. [1] the natural units of measurement ($\hbar = c = 1$) is applied), e.g., for guided photons inside the hollow metallic waveguide being placed along the direction of x-axis, its dispersion relation $(\hbar\omega)^2 = (\hbar k_x)^2 c^2 + (\hbar\omega_c/c^2)^2 c^4$ is exactly similar to Einstein's relation $E^2 = p^2c^2 + m^2c^4$ for a particle (with the rest mass m, energy E and momentum p), where $\omega$ is the photons' frequency and $k_x$ is the x-component of photons' wavenumber vector, $\omega_c$ is the lowest-order cutoff frequency of the waveguide (for simplicity our discussion is restricted to the lowest-order cutoff frequency). As a result, $m_{\text{eff}} = \hbar\omega_c/c^2$ plays the role of effective rest mass of photons inside the waveguide.

Now, let such waveguide rest in the classical reference frame $S$, and let the particle Q discussed above be identical with such guided photons. In the present case, as observed in the classical reference frame $S$, the guided photons inside the waveguide has the dispersion relation $E_p^2 = (\hbar\omega)^2 = \boldsymbol{p}_L^2 c^2 + m_{\text{eff}}^2 c^4$, and has the four-dimensional (4D) momentum $p_L^\mu = (E_p, c\boldsymbol{p}_L) = (\hbar\omega, c\hbar k_x, 0, 0)$, where $E_p = \hbar\omega$ represents the energy of the guided photons, $\boldsymbol{p}_L = (\hbar k_x, 0, 0)$ represents the 3D momentum of the guided photons propagating along the direction of the waveguide, and $m_{\text{eff}} = \hbar\omega_c/c^2$ represents the effective rest mass of the guided photons. Furthermore, the quantum reference frame $S'$ is attached to the guided photons with the effective rest mass $m_{\text{eff}} = \hbar\omega_c/c^2$, such that the guided photons' group velocity $\boldsymbol{v}_g = (c^2 k_x/\omega, 0, 0,)$ along the waveguide is the relative velocity between the frames $S$ and $S'$. In terms of the effective rest mass $m_{\text{eff}} = \hbar\omega_c/c^2$, the effective Compton wavelength of the guided photons is defined as

$$\lambdabar_c \equiv \hbar/m_{\text{eff}} c = \hbar c/\hbar\omega_c, \tag{14}$$

Likewise, according to Section **3**, the guided photons can propagate over a space-like interval provided that

$$0 > c^2 t^2 - x^2 \geq -(\lambdabar_c/2)^2, \tag{15}$$

On the other hand, let $\Gamma_p(t, x)$ denote the probability amplitude for the guided photons to propagate from $(0,0)$ to $(t, x)$ along the waveguide. According to Refs. [1-3], for spacelike interval $c^2t^2 - x^2 < 0$ (for the moment the guided photons tunneling through an undersized waveguide), one has:

$$\Gamma_p(t,x) \propto \exp(-\omega_c \sqrt{x^2 - c^2t^2}/c) = \exp(-\sqrt{x^2 - c^2t^2}/\lambdabar_c). \tag{16}$$

Therefore, as for the probability (say, $P_p(t,x)$) for the guided photons to propagate from $(0,0)$ to $(t,x)$ inside the undersized waveguide, one has

$$P_p(t,x) \propto |\Gamma_p(t,x)|^2 \propto \exp(-2\sqrt{x^2 - c^2t^2}/\lambdabar_c). \tag{17}$$

Therefore, as $\exp(-2\sqrt{x^2 - c^2t^2}/\lambdabar_c) \geq \exp(-1) \approx 0$, i.e., $0 > c^2t^2 - x^2 \geq -(\lambdabar_c/2)^2$, the probability $P_p(t,x)$ for photons inside an undersized waveguide to superluminally propagate from $(0,0)$ to $(t,x)$ cannot be ignored. Likewise, Eq. (15) is obtained at the quantum-mechanical level while Eq. (17) is obtained at quantum-field-theory level. As a result, Eq. (15) can be obtained by taking the numerical approximation of Eq. (17), i.e., the conventional numerical approximation $\exp(-2\sqrt{x^2 - c^2t^2}/\lambdabar_c) \geq \exp(-1) \approx 0$.

**5. Conclusions and discussions**

As far as the superluminality of evanescent modes tunneling through an undersized waveguide is concerned, the conclusion from quantum Lorentz transformation accords with that from quantum field theory. As a purely quantum-mechanical effect, the presence of the term $\hbar^2 c^2 \hat{H}^{-2}/4$ in Eq. (9) is essentially due to the commutation relation $[x,\hat{p}] = i\hbar$. Therefore, the fact that a particle with finite mass can propagate over a spacelike interval attributes to the Heisenberg's uncertainty relation. Such a superluminal phenomenon preserves a quantum-mechanical causality [1-2].

Note that in our case, spacetime coordinates are also spacetime intervals (with respect to origins of coordinates). As mentioned before, as viewed in the classical reference frame $S$ one has $xt = tx$, and time enters as a parameter rather than an operator. On the other hand, one can prove that (see **Appendix B**):

$$x't' - t'x' = -i\hbar(\hat{H}^{-1}x + x\hat{H}^{-1})/2. \tag{18}$$

That is, as viewed in the quantum reference frame $S'$, the spacetime coordinates of the particle Q are noncommutative and time enters as an operator. In fact, once time enters as an operator, spacetime coordinates may become noncommutative. For example, let $p_x = mu$, by quantizing the classical expression $t = \pm x/u = \pm mx/p$ one can obtain the nonrelativistic free-motion time-of-arrival operator $\hat{T}_{non} = \pm m(\hat{p}^{-1}x + x\hat{p}^{-1})/2$ [34-37]. If we take $\hat{T}_{non} = m(\hat{p}^{-1}x + x\hat{p}^{-1})/2$, and note that in the momentum space representation one has $\hat{x} = i\hbar\partial/\partial p$ and $\hat{x}\hat{p}^{-1} - \hat{p}^{-1}\hat{x} = -i\hbar\hat{p}^{-2}$, one can prove that

$$x\hat{T}_{non} - \hat{T}_{non}x = -i\hbar(\hat{H}_{non}^{-1}x + x\hat{H}_{non}^{-1})/4. \tag{19}$$

where $\hat{H}_{non} = \hat{p}^2/2m$. Eq. (19) implies that there is an uncertainty relation between the time-of-arrival and position-of-arrival.

**Acknowledgments**

This work was supported by the National Natural Science Foundation of China (Grant No. 60671030).

**Appendix A: Proof of Eq. (9)**

Using Eq. (5) one has

$$c^2 t'^2 - x'^2 = [\frac{t\hat{H}}{mc} - \frac{(\hat{p}x + x\hat{p})}{2mc}]^2 - [\frac{(\hat{H}x + x\hat{H})}{2mc^2} - \frac{t\hat{p}}{m}]^2, \tag{a1}$$

Using $\hat{H}t = t\hat{H}$, $t\hat{p} = \hat{p}t$, $xt = tx$ and $\hat{H}\hat{p} = \hat{p}\hat{H}$ one has

$$\frac{1}{2m^2c^2}[-t\hat{H}(\hat{p}x + x\hat{p}) - (\hat{p}x + x\hat{p})t\hat{H} + (\hat{H}x + x\hat{H})t\hat{p} + t\hat{p}(\hat{H}x + x\hat{H})] = 0, \tag{a2}$$

then

$$c^2 t'^2 - x'^2 = \frac{t^2 \hat{H}^2}{m^2 c^2} + \frac{(\hat{p}x + x\hat{p})(\hat{p}x + x\hat{p})}{4m^2 c^2} - \frac{(\hat{H}x + x\hat{H})(\hat{H}x + x\hat{H})}{4m^2 c^4} - \frac{t^2 \hat{p}^2}{m^2}. \tag{a3}$$

Using $\hat{p}x = x\hat{p} - i\hbar$, $\hat{H}x = x\hat{H} - i\hbar\hat{H}^{-1}\hat{p}c^2$, and $\hat{H}^2 = \hat{p}^2 c^2 + m^2 c^4$, one has

$$c^2 t'^2 - x'^2 = c^2 t^2 + \frac{(2x\hat{p} - i\hbar)(2x\hat{p} - i\hbar)}{4m^2 c^2} - \frac{(2x\hat{H} - i\hbar\hat{H}^{-1}\hat{p}c^2)(2x\hat{H} - i\hbar\hat{H}^{-1}\hat{p}c^2)}{4m^2 c^4}. \tag{a4}$$

Because

$$\begin{aligned}(2x\hat{p} - i\hbar)(2x\hat{p} - i\hbar) &= 4x\hat{p}x\hat{p} - 4i\hbar x\hat{p} - \hbar^2 \\ &= 4x(x\hat{p} - i\hbar)\hat{p} - 4i\hbar x\hat{p} - \hbar^2 = 4x^2\hat{p}^2 - 8i\hbar x\hat{p} - \hbar^2\end{aligned}, \tag{a5}$$

$$\begin{aligned}&(2x\hat{H} - i\hbar\hat{H}^{-1}\hat{p}c^2)(2x\hat{H} - i\hbar\hat{H}^{-1}\hat{p}c^2) \\ &= 4x\hat{H}x\hat{H} - 2i\hbar x\hat{p}c^2 - 2i\hbar\hat{p}c^2\hat{H}^{-1}x\hat{H} - \hbar^2\hat{H}^{-2}\hat{p}^2 c^4 \\ &= 4x(x\hat{H} - i\hbar\hat{H}^{-1}\hat{p}c^2)\hat{H} - 2i\hbar x\hat{p}c^2 - 2i\hbar\hat{p}c^2\hat{H}^{-1}(\hat{H}x + i\hbar\hat{H}^{-1}\hat{p}c^2) - \hbar^2\hat{H}^{-2}\hat{p}^2 c^4 \\ &= 4x^2\hat{H}^2 - 4i\hbar x\hat{p}c^2 - 2i\hbar x\hat{p}c^2 - 2i\hbar\hat{p}c^2 x + 2\hbar^2\hat{H}^{-2}\hat{p}^2 c^4 - \hbar^2\hat{H}^{-2}\hat{p}^2 c^4 \\ &= 4x^2\hat{H}^2 - 6i\hbar x\hat{p}c^2 - 2i\hbar c^2(x\hat{p} - i\hbar) + \hbar^2\hat{H}^{-2}\hat{p}^2 c^4 \\ &= 4x^2\hat{H}^2 - 8i\hbar x\hat{p}c^2 - 2\hbar^2 c^2 + \hbar^2\hat{H}^{-2}\hat{p}^2 c^4\end{aligned}, \tag{a6}$$

one has

$$\frac{(2x\hat{p}-i\hbar)(2x\hat{p}-i\hbar)}{4m^2c^2} - \frac{(2x\hat{H}-i\hbar\hat{H}^{-1}\hat{p}c^2)(2x\hat{H}-i\hbar\hat{H}^{-1}\hat{p}c^2)}{4m^2c^4}$$

$$= \frac{1}{4m^2c^4}[(4x^2\hat{p}^2c^2 - 8i\hbar x\hat{p}c^2 - \hbar^2c^2) - (4x^2\hat{H}^2 - 8i\hbar x\hat{p}c^2 - 2\hbar^2c^2 + \hbar^2\hat{H}^{-2}\hat{p}^2c^4)]$$

$$= \frac{1}{4m^2c^4}[-4x^2m^2c^4 + \hbar^2c^2 - \hbar^2\hat{H}^{-2}\hat{p}^2c^4] \qquad , \qquad \text{(a7)}$$

$$= -x^2 + \frac{1}{4m^2c^4}[\hbar^2c^2\hat{H}^{-2}(\hat{H}^2 - \hat{p}^2c^2)]$$

$$= -x^2 + \hbar^2c^2\hat{H}^{-2}/4$$

then one has

$$c^2t'^2 - x'^2 = c^2t^2 - x^2 + \hbar^2c^2\hat{H}^{-2}/4, \qquad \text{(a8)}$$

which is exactly Eq. (9).

**Appendix B: Proof of Eq. (18)**

Using $\hat{H}t = t\hat{H}$, $t\hat{p} = \hat{p}t$, $xt = tx$, $\hat{H}\hat{p} = \hat{p}\hat{H}$ and Eq. (5), one has

$$x't' - t'x' = \frac{t}{2m^2c^4}\{[(\hat{H}x + x\hat{H})\hat{H} - \hat{H}(\hat{H}x + x\hat{H})] + c^2[\hat{p}(\hat{p}x + x\hat{p}) - (\hat{p}x + x\hat{p})\hat{p}]\}$$

$$+ \frac{(\hat{p}x + x\hat{p})(\hat{H}x + x\hat{H}) - (\hat{H}x + x\hat{H})(\hat{p}x + x\hat{p})}{4m^2c^4} \qquad . \qquad \text{(b1)}$$

Using $\hat{p}x = x\hat{p} - i\hbar$, $\hat{H}x = x\hat{H} - i\hbar\hat{H}^{-1}\hat{p}c^2$, and $\hat{H}^2 = \hat{p}^2c^2 + m^2c^4$, one has

$$[(\hat{H}x + x\hat{H})\hat{H} - \hat{H}(\hat{H}x + x\hat{H})] + c^2[\hat{p}(\hat{p}x + x\hat{p}) - (\hat{p}x + x\hat{p})\hat{p}]$$

$$= [x\hat{H}^2 - \hat{H}^2x] + c^2[\hat{p}^2x - x\hat{p}^2]$$

$$= x(\hat{H}^2 - \hat{p}^2c^2) + (\hat{p}^2c^2 - \hat{H}^2)x \qquad , \qquad \text{(b2)}$$

$$= xm^2c^4 - m^2c^4x = 0$$

then

$$4m^2c^4(x't' - t'x') = (\hat{p}x + x\hat{p})(\hat{H}x + x\hat{H}) - (\hat{H}x + x\hat{H})(\hat{p}x + x\hat{p})$$

$$= (2x\hat{p} - i\hbar)(2\hat{H}x + i\hbar c^2\hat{H}^{-1}\hat{p}) - (2x\hat{H} - i\hbar c^2\hat{H}^{-1}\hat{p})(2\hat{p}x + i\hbar)$$

$$= 4x\hat{p}\hat{H}x + 2x\hat{p}i\hbar c^2\hat{H}^{-1}\hat{p} - i\hbar 2\hat{H}x + \hbar^2c^2\hat{H}^{-1}\hat{p} - 4x\hat{H}\hat{p}x$$

$$- i\hbar 2x\hat{H} + i\hbar c^2\hat{H}^{-1}\hat{p}2\hat{p}x - \hbar^2c^2\hat{H}^{-1}\hat{p} \qquad . \qquad \text{(b3)}$$

$$= 2i\hbar x\hat{H}^{-1}c^2\hat{p}^2 - 2i\hbar x\hat{H} + c^2\hat{p}^2\hat{H}^{-1}2i\hbar x - \hat{H}2i\hbar x$$

$$= 2i\hbar x\hat{H}^{-1}(c^2\hat{p}^2 - \hat{H}^2) + (c^2\hat{p}^2 - \hat{H}^2)2i\hbar\hat{H}^{-1}x$$

$$= -2m^2c^4i\hbar(x\hat{H}^{-1} + \hat{H}^{-1}x)$$

It follows that

$$x't' - t'x' = -i\hbar(\hat{H}^{-1}x + x\hat{H}^{-1})/2. \tag{b4}$$

Then we obtain Eq. (18).